\documentclass[aps,prb,twocolumn,superscriptaddress,longbibliography]{revtex4-1}

\usepackage[colorlinks=true,citecolor=blue]{hyperref}
\usepackage{graphicx}
\usepackage{amsmath}
\usepackage{amssymb}

\DeclareGraphicsExtensions{.png,.jpg,.eps}
\usepackage{xcolor}

\renewcommand{\H} {\mathcal{H}}

\begin{document}

\author{D. Soriano}
\affiliation{Radboud University, Institute for Molecules and Materials, NL-6525 AJ Nijmegen, the Netherlands}
\affiliation{Dipartimento di Ingegneria dell'Informazione, Universitá di Pisa, 56122 Pisa, Italy}

\author{J. L. Lado}
\affiliation{Department of Applied Physics, Aalto University, 00076 Aalto, Espoo, Finland}

\title{
    Spin-orbit
    correlations
    and exchange-bias control
     in twisted Janus dichalcogenide multilayers
}

\begin{abstract}
Janus dichalcogenide multilayers provide a paradigmatic platform to engineer electronic phenomena dominated by spin-orbit coupling. Their unique spin-orbit effects stem from the mirror symmetry breaking in each layer, which induces a colossal Rashba spin-orbit effect in comparison with the conventional dichalcogenide counterparts. Here we put forward twisted dichalcogenide bilayers as a simple platform to realize spin-orbit correlated states. We demonstrate the emergence of flat bands featuring strong spin-momentum locking and the emergence of symmetry broken states with associated non-coplanar magnetization when interactions are included.  We further show that the symmetry broken states can be controlled by means of a magnetic substrate, strongly impacting the non-coplanar magnetic texture of the moire unit cell.  Our results put forward twisted Janus multilayers as a powerful platform to explore spin-orbit correlated physics, and highlighting the versatility of magnetic substrates to control unconventional moire magnetism.
\end{abstract}

\date{\today}

\maketitle

\section{Introduction}

Dichalcogenide van der Waals materials \cite{Radisavljevic2011,Splendiani2010,Manzeli2017,PhysRevLett.105.136805,Dickinson1923,Wilson1969} are paradigmatic two dimensional compounds featuring strong spin-orbit coupling effects\cite{PhysRevX.4.011034,Mak2012,Zeng2012,Cao2012,PhysRevB.86.081301,Xu2014,Mak2016}.  The strong spin-momentum locking has motivated a variety of proposals for spin-orbitronic control. \cite{PhysRevLett.119.196801,Safeer2019,Shao2016,Mendes2018,Luo2017,Avsar2017} These phenomena comes from the intrinsic Ising spin-orbit coupling, stemming from the inversion symmetry breaking of the structure.  Interestingly, a special family of dichalcogenides, known as Janus dichalcogenides\cite{Lu2017,Dong2017,Li2017,PhysRevB.100.165425}, provide a unique new form of spin-orbit physics to these materials. Janus dichalcogenides feature an intrinsic mirror-symmetry breaking adding strong Rashba spin-orbit coupling effects\cite{Sant2020,2020arXiv200707579S,PhysRevB.103.035414} to the already rich physics of conventional dichalcogenides.

Twist engineering has risen as a powerful knob to control the electronic structure of dichalcogenides in particular,\cite{PhysRevLett.122.086402,PhysRevLett.121.266401,PhysRevResearch.2.013335,PhysRevB.91.165403,Jones2013,Baugher2014} and van der Waals materials in general.\cite{PhysRevB.86.155449,PhysRevLett.99.256802,PhysRevLett.124.086401,Chen2020mo,Hejazi2020} The emergence of flat bands and correlated states in dichalcogenide systems represent an illustrative example of the versatility of twist engineering in these systems.  \cite{PhysRevLett.121.026402,PhysRevLett.126.106804,Jin2019TM,Regan2020TM,Shimazaki2020,Shimazaki2020,Tang2020TM,PhysRevB.103.L121102} Spin-orbit coupling effects in these flat band systems\cite{Hill2016,PhysRevB.92.205108,PhysRevB.84.153402,PhysRevLett.108.196802,Soriano2020,PhysRevB.102.241106,2021arXiv210203259V} have also been proposed to bring up unique possibilities for spin and valley control. However, the emergence of correlated states in Janus dichalcogenides has remained largely unexplored, and studies on spin-orbit effects have focused mainly on Ising spin-orbit coupling.

\begin{figure}[t!]
\centering
    \includegraphics[width=\columnwidth]{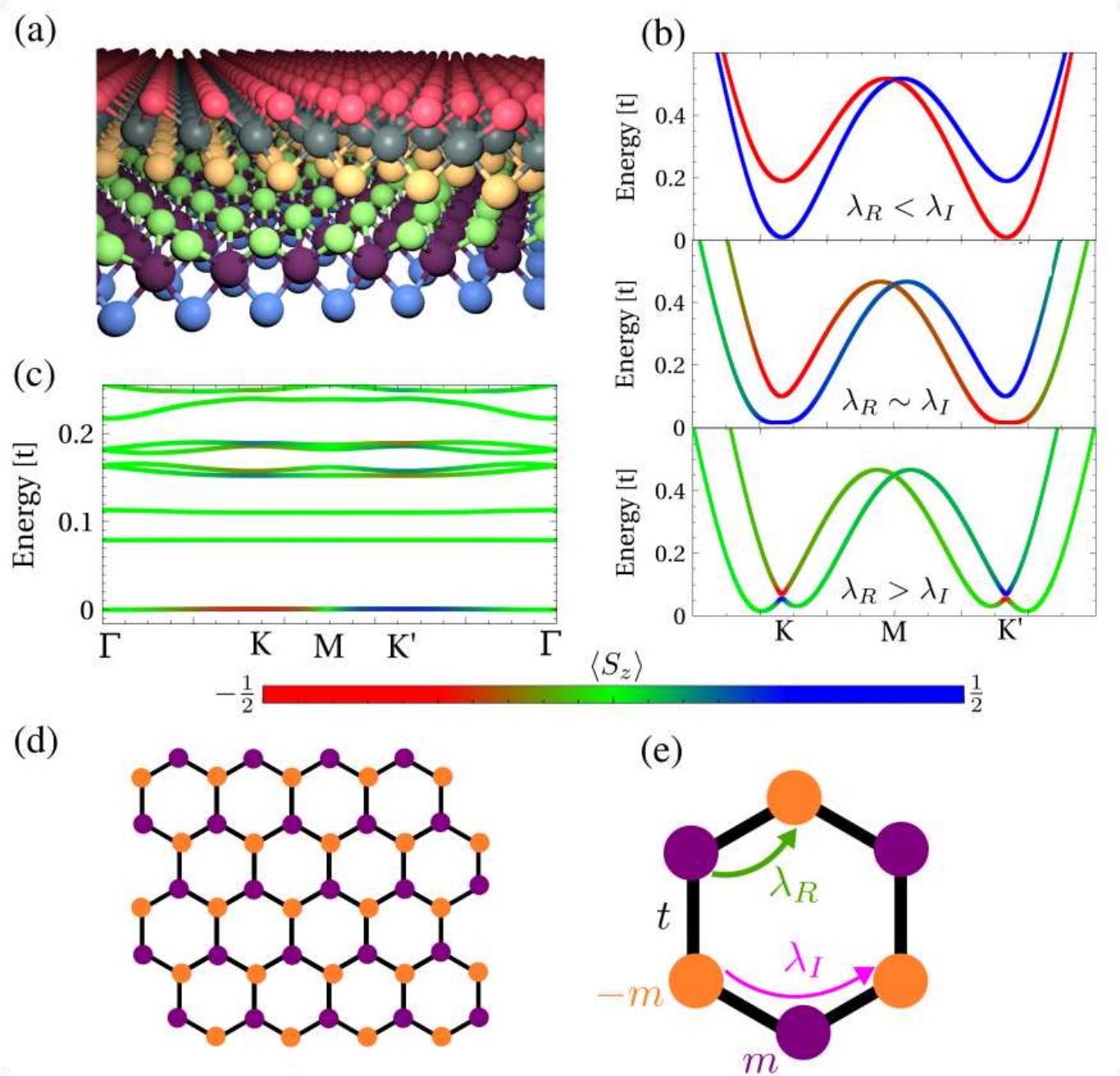}

\caption{
    (a) Sketch of a twisted Janus dichalcogenide, showing in different
    colors the different chalcogenide atoms.
    Panel (b) shows the low energy band structure of
    a monolayer Janus dichalcogenide as the strength of the
    Rashba spin-orbit $\lambda_R$ coupling is increased
    in comparison with the Ising spin-orbit coupling $\lambda_I$.
    Panel (c) shows the moire bands of twisted Janus system
    at a rotation angle of 2$^\circ$, at the bottom of the conduction band.
    Panel (d) shows an illustration of the low energy model, and panel
    (e) shows the different terms in the Hamiltonian.
    We took $\lambda_I=0.02t$ in (b,c). 
}
\label{fig:fig1}
\end{figure}

Here we put forward twisted Janus systems (Fig. \ref{fig:fig1}a) as a paradigmatic example in which correlated states emerge in flat bands dominated by spin-momentum texture in reciprocal space.  In particular, we demonstrate the emergence of flat bands dominated by Rashba spin-orbit coupling effects in strike contrast with conventional twisted dichalcogenide systems.  When including electronic interactions, we show that these flat bands lead to non-coplanar magnetic states in real space, stemming from the strong Rashba effects of the parent compounds.  We finally show how the exchange proximity effect allows controlling these correlated states by combining Janus dichalcogenide bilayers with two-dimensional ferromagnets.  Our results highlight the spin-orbit dominated physics of twisted Janus dichalcogenide systems and their versatility for controllable correlated spin-orbitronics.  The manuscript is organized as follows.  In Section \ref{sec:bands} we show the impact of mirror symmetry breaking in twisted dichalcogenide systems.  In Section \ref{sec:int} we show the emergence of correlations in the twisted Janus system, together with the emergence of non-coplanar magnetism.  In Section \ref{sec:ex} we show how the correlated states can be controlled by the exchange proximity effect.  Finally, in Section \ref{sec:con} we summarize our conclusions.

\begin{figure}[t!]
\centering
    \includegraphics[width=\columnwidth]{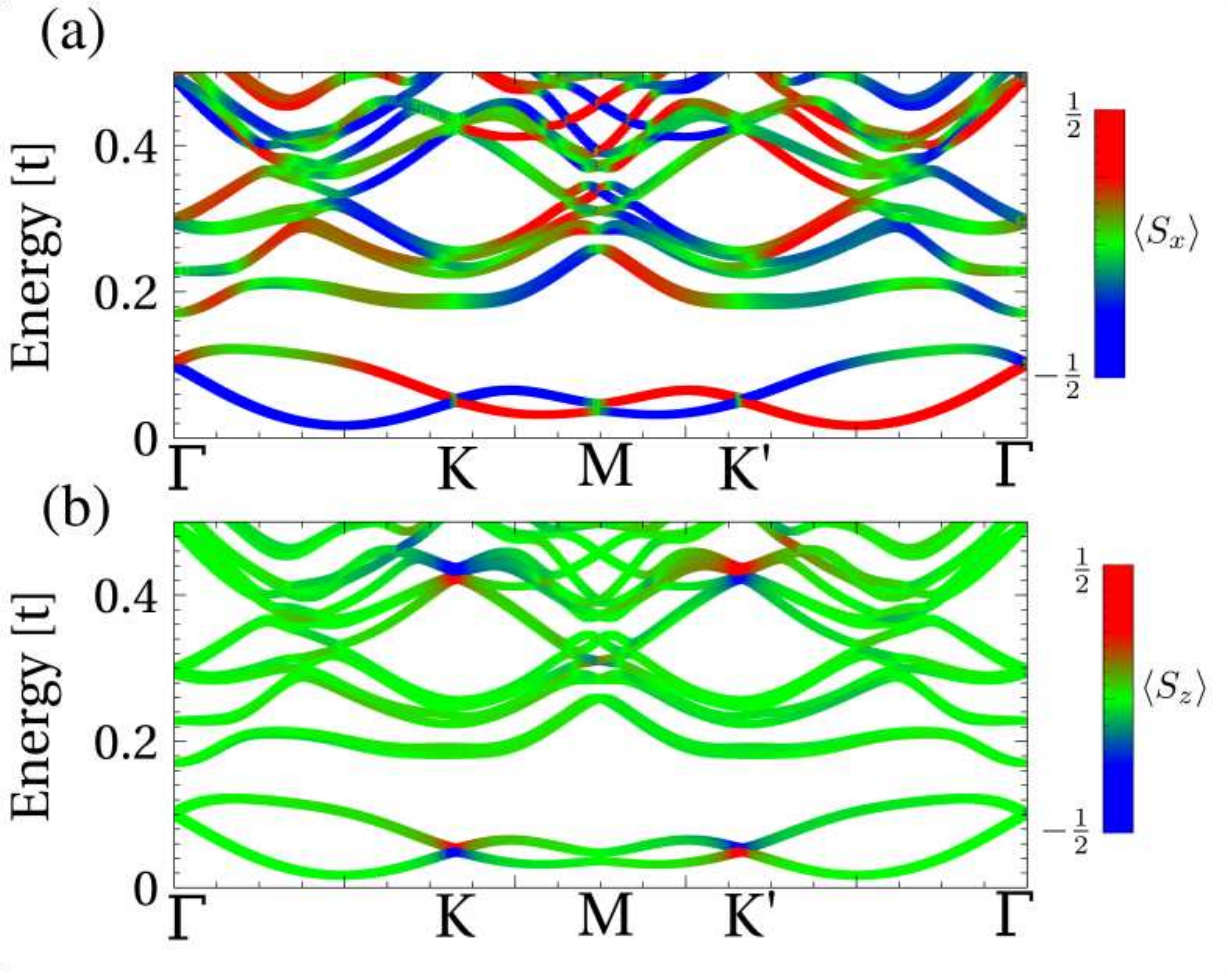}

\caption{
	(a,b) Band structure of a twisted Janus
    dichalcogenide at a twist angle of $9^\circ$,
    showing the $S_x$ (a) and $S_z$ (b) 
    spin textures in reciprocal space.
    The strong spin-momentum texture arises from the
    intrinsic mirror symmetry
    breaking of each Janus monolayer,
    that is inherited by the twisted bilayer
    system.
    We took $\lambda_I = 0.1 \lambda_R$ and $\lambda_I=0.02t$ in (a,b).
}
\label{fig:fig2}
\end{figure}

\section{Spin-texture in twisted Janus bilayers}
\label{sec:bands}
We first address the emergence of flat bands in the Janus van der Waals heterostructure, and in particular, how the different types of spin-orbit affect the low energy electronic structure. 
In particular, Janus materials have coexisting types of spin-orbit coupling, namely Ising and Rashba \cite{Sant2020,2020arXiv200707579S,PhysRevB.103.035414}, whose relative strength substantially
impact the low energy electronic dispersion (Fig. \ref{fig:fig1}b). Such interplay of spin-orbit effects further impacts the
quasi flat bands of a twisted heterostructure (Fig. \ref{fig:fig1}c), and therefore would be expected to have an importan
impact in potentially correlated states. To model these systems, we consider an effective real-space model for the Janus material that incorporates the effects of intrinsic Ising spin-orbit coupling and Rashba spin-orbit coupling.  We will focus the discussion on the conduction band of the model, yet an analogous discussion can be performed for the valence band.  The Janus dichalcogenide system is captured by an effective honeycomb lattice model (Fig. \ref{fig:fig1}d) of the following form

\begin{multline}
	\mathcal{H} = \sum_{\langle ij \rangle,s} t c^\dagger_{i,s} c_{j,s}
	+ m\sum_{i,s} \vartheta^z_{ii} c^\dagger_{i,s} c_{i,s}  \\
	+ \sum_{i,j,s} 
    t^\perp (\mathbf{r}_i,\mathbf{r}_j) c^\dagger_{i,s} c_{j,s} \\
    + i\lambda_{I} \sum_{\langle\langle ij \rangle\rangle,s,s'} 
	\nu_{ij}
	\sigma^z_{s,s'} c^\dagger_{i,s} c_{j,s'} \\
    + i\lambda_R \sum_{\langle ij \rangle,s,s'} \mathbf{d}_{ij} \cdot
    \mathbf {\sigma}_{s,s'} c^\dagger_{i,s} c_{j,s'} \\
    \label{eq:h0}
\end{multline}

where $t$ is a first neighbor hopping, $\langle \rangle$ denotes sum over first neighbors in a layer, $m$ denotes the onsite energy imbalance between the two sites of the honeycomb lattice, $t^\perp (\mathbf{r}_i,\mathbf{r}_j)$ parametrize the interlayer coupling,\cite{PhysRevB.82.121407,PhysRevB.92.075402,PhysRevLett.119.107201} $\vartheta^z_{ii}$ is the sublattice Pauli matrix, $\sigma^z_{s,s'}$ is the spin Pauli matrix, $\lambda_I$ controls the Ising spin-orbit coupling $\nu_{ij}=\pm 1$ for clock-wise/anti-clockwise hopping,\cite{PhysRevLett.95.226801}, $\langle \langle \rangle \rangle$ denotes second neighbors in a layer, $\lambda_R$ controls the Rashba spin-orbit coupling\cite{PhysRevB.82.161414} and
$\mathbf{d}_{ij} = (\mathbf{r}_i - \mathbf{r}_j)\times \hat z$. An illustration
of the different terms of the effective model of a Janus monolayer is shown in Fig. \ref{fig:fig1}e.
For a conventional dichalcogenide system, i.e., non-Janus one, $\lambda_R=0$ due to the presence of mirror symmetry in the layer, and the Hamiltonian has $U(1)$ rotational spin symmetry.  For a Janus system, $\lambda_R\ne 0$ breaks the $U(1)$ rotational spin-symmetry, creating a spin-momentum texture in reciprocal space.  For the sake of simplicity, we take the same sign of $\lambda_R$ for both layers, which physically means the two systems have the same mirror symmetry breaking. The values of $\lambda_I$ and $\lambda_R$ are material specific, and can be controlled
by choosing different dichalcogenide semiconductors\cite{PhysRevB.88.245436,PhysRevB.95.165401}.
From the electronic structure point of view, the value of $\lambda_I$ controls the splitting at the K-point, whereas
the value of $\lambda_R$ controls the momentum offset of the two spin channels.
As a reference, $\lambda_R$ can reach up to 50 meV in WSeTe\cite{PhysRevB.95.165401}, and
$\lambda_I$ can range between 1 and 100 meV in different 
dichalcogenides\cite{PhysRevB.88.245436}.
Such parameters can be extracted from the spin-splitting of the electronic structure
computed with first principle methods including spin-orbit coupling effects\cite{PhysRevB.95.165401,PhysRevB.88.245436}.
In the single-layer limit, the model of Eq. \ref{eq:h0} captures the dispersion of a Janus monolayer in the presence of spin-orbit coupling.  We finally note that this model captures mainly the electronic structure at low energies and that multiorbital Wannier models\cite{Pizzi2020} could be used to capture the physics at higher energies.

We now move on to discuss in more detail the electronic structure of the twisted Janus system. In the presence of a twist, the moire length $L_M$ emerges and can be controlled by the twist angle, and the electronic structure now is defined in the mini-Brillouin zone. The strong spin-momentum locking of the monolayer system is inherited by the electronic structure of the twisted Janus system, as shown in Fig. \ref{fig:fig2}.  In particular, we first take a twisted system at a large twist angle of $9^\circ$ degrees and compute the spin expectation value of the $S_x$ operator and $S_z$ operator in reciprocal space.  As shown in Fig. \ref{fig:fig2}ab, we observe a strong texture of the $S_x$ operator in reciprocal space (Fig. \ref{fig:fig2}a), with the $S_z$ operator strongly mixed (Fig. \ref{fig:fig2}b).  This behavior is in stark contrast with non-Janus systems, in which the Ising spin-orbit coupling creates spin-polarized moire bands in reciprocal space.

\begin{figure}[t!]
\centering
    \includegraphics[width=\columnwidth]{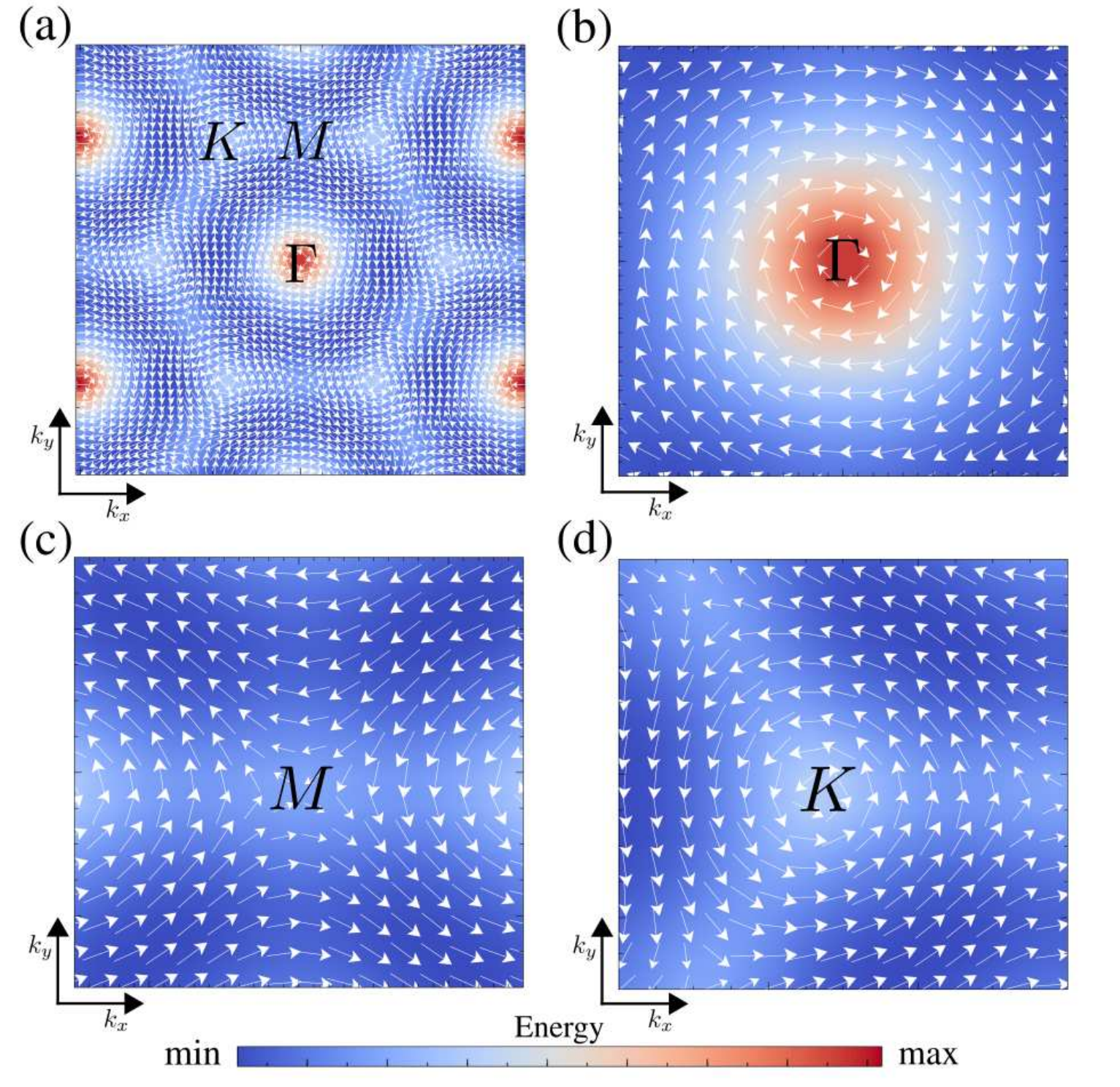}

\caption{ (a) Spin texture in momentum
space of the lowest conduction band
of a tiwsted Janus bilayer at an angle of
9$^\circ$. Vortices around the 
$\Gamma$ (b), $M$ (c) and $K$ (d) are clearly observed.
The arrows denote the expectation value
$(\langle S_x \rangle, \langle S_y \rangle)$
for that eigenstate.
 We took $\lambda_I = 0.1 \lambda_R$ and $\lambda_I=0.02t$.
}
\label{fig:fig3}
\end{figure}

The spin-momentum texture of the Janus system can be further explored by computing the spin projection in the lowest conduction band. In particular, we show in Fig. \ref{fig:fig3} a quiver map of the lowest conduction band, where the arrows denote the expectation value of the spin operators $S_x$ and $S_y$, and the color the eigenvalue of the state. A strong spin texture in the whole reciprocal space is clearly seen (Fig. \ref{fig:fig3}a). In particular, spin texture vortices are observed around the $\Gamma$ (Fig. \ref{fig:fig3}b), $M$ (Fig. \ref{fig:fig3}c) and $K$ (Fig. \ref{fig:fig3}d) points. This in-plane spin texture is absent in conventional twisted dichalcogenide multilayers and therefore is a unique feature of twisted Janus bilayers.  This spin texture will heavily influence the correlated states in the system, as we discuss in the next section.

\section{Interaction-induced symmetry breaking in 
Janus systems}
\label{sec:int}

After having addressed the single-particle electronic structure, we now move on to consider the effect of electron-electron interactions,
focusing on a system with a smaller twist angle and flatter bands. We include non-local interactions up second neighbor sites in our atomistic model, taking the form

\begin{multline}
	\H_I = 
	U
	\sum_i 
    \left ( c^\dagger_{i,\uparrow}c_{i,\uparrow} -\frac{1}{2}\right )
	\left ( c^\dagger_{i,\downarrow}c_{i,\downarrow} - \frac{1}{2} \right ) \\
	+ V_1
	\sum_{\langle ij \rangle} \left [
	\left ( \sum_s c^\dagger_{i,s}c_{i,s}  \right )
	\left ( \sum_s c^\dagger_{j,s}c_{j,s}  \right )
	\right ]
    \\
	+ V_2
	\sum_{\langle \langle ij \rangle \rangle} \left [
	\left ( \sum_s c^\dagger_{i,s}c_{i,s}  \right )
	\left ( \sum_s c^\dagger_{j,s}c_{j,s}  \right )
	\right ]
    \\
\end{multline}

where $U$ parametrizes the local Hubbard interaction, $V_1$ the first neighbor interaction and $V_2$ the second neighbor interaction. By solving the previous Hamiltonian via a mean-field approximation, we can address the interaction-induced symmetry broken states in the Janus system.  Importantly, we will exploit a fully rotational invariant mean-field formalism capable of capturing non-coplanar formalism. Our mean-field formalism includes all the Wick contractions in the mean-field ansatz, which in particular captures interaction-induced real-space non-coplanar moire spin-textures.  For the sake of concreteness, we will focus on correlated states in the case with two electrons per unit cell.

\begin{figure}[t!]
\centering
    \includegraphics[width=\columnwidth]{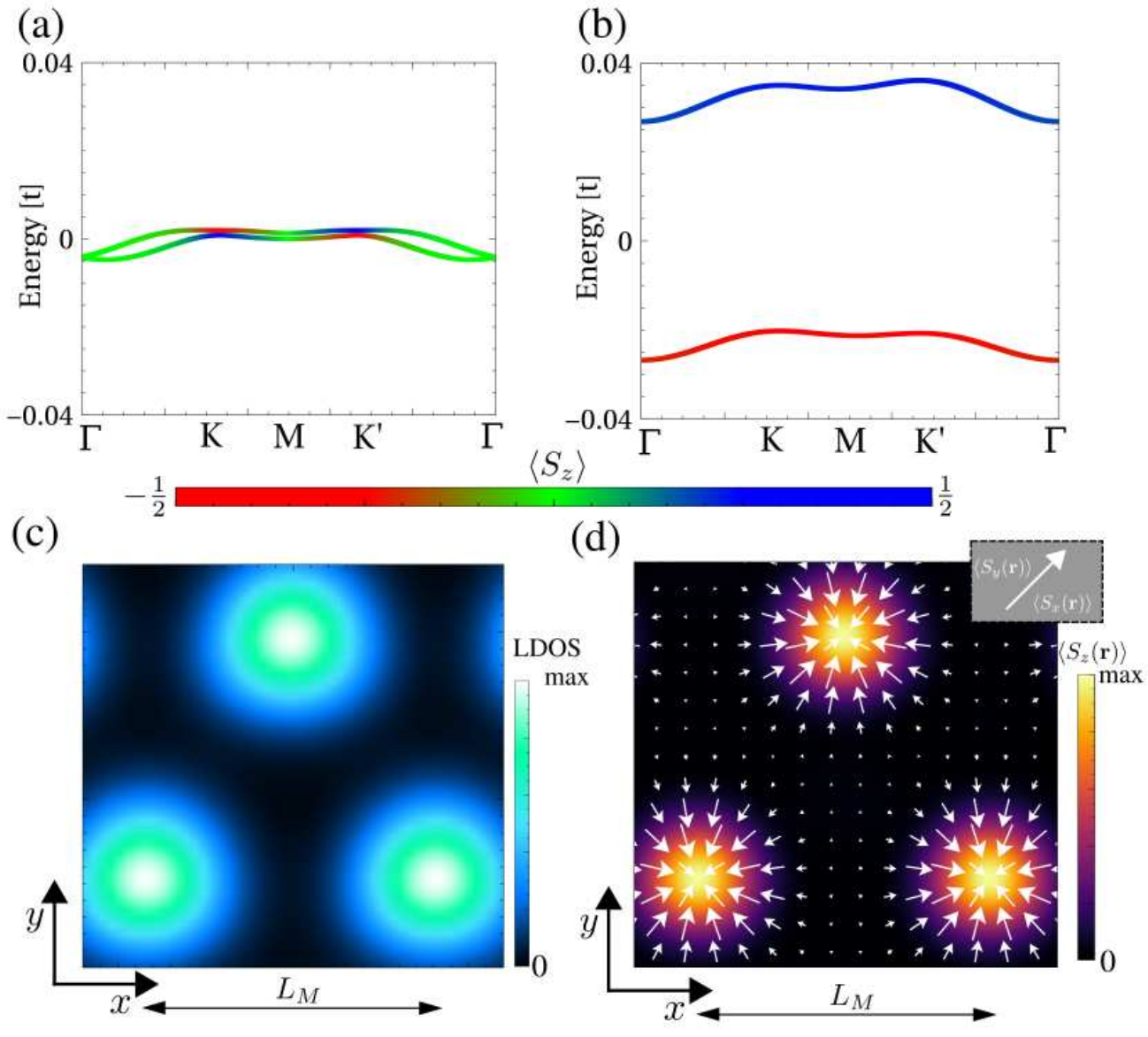}

\caption{ (a) Non-interacting
and (b) selfconsistent
electronic structure of the twisted
Janus dichalcogenide
at a twist angle of 3.5$^\circ$.
Before interactions, the
nearly flat bands form a
effective triangular
lattice in real-space (c).
The interaction-induced magnetization
displays a strong texture in real-space
as shown in panel (d),
stemming from the Rashba spin-orbit coupling
of the twisted dichalcogenide.
$L_M$ denotes the moire length.
We took $U=1.5t$, $V_1 = 1.2t$ and $V_2 = 0.3t$
and $\lambda_I = 0.1 \lambda_R$ and $\lambda_I=0.02t$.  
}
\label{fig:fig4}
\end{figure}

We now address the impact of interactions in the twisted Janus system.  We show in Fig. \ref{fig:fig4}a the low energy electronic structure for a twisting angle of 3.5$^\circ$, featuring four nearly flat bands with strong spin-momentum texture.  Once we include electronic interactions at the mean-field level, we observe the emergence of a ferromagnetic state at half-filling of the low energy bands, as shown in Fig. \ref{fig:fig4}b.  The low energy bands in the absence of interactions form a triangular lattice in the moire supercell, as shown in the density of states of Fig. \ref{fig:fig4}c. The real-space spin density in the correlated regime also presents a non-zero expectation value in the emergent moire triangular lattice, as shown in Fig. \ref{fig:fig4}d.  This phenomenology is, of course, common to most of the moire systems featuring strongly localized modes in real space. However, more interestingly, the moire system displays a strong non-coplanar magnetization in real space, forming a meron texture around each moire hot spot. 
It is worth to note that the local order parameter is not only non-collinear,
but also that is not contained in a plane and thus we refer to it as non-coplanar.
The emergence of such non-coplanar real-space spin texture stems from the intrinsic mirror symmetry breaking of each Janus monolayer, and in particular, from the intrinsic strong Rashba spin-orbit coupling.

We now comment on some observable transport signatures associated to this correlated state.
The existence of a non-coplanar magnetic texture is expected to have
an important impact on the electronic transport properties of the system. Upon
small doping away from the correlated state, free electrons will feel a strong non-coplanar
spin-texture in real space that will lead to anomalous Hall effect\cite{RevModPhys.82.1959}. The emergence
of this anomalous Hall state is well-known in other non-coplanar magnetic systems, with
the paradigmatic case of skyrmion materials\cite{Nagaosa2013,PhysRevB.80.054416,Kurumaji2019}. In the strong coupling limit, by performing a local
gauge rotation to the local spin axis, the effective equations of motion for electrons
acquire an effective gauge field associated with the non-coplanar magnetization\cite{Nagaosa2013}. This new term enters
as an anomalous velocity in the equations of motion, and it is known to give rise to the so-called
anomalous skyrmion Hall effect\cite{Nagaosa2013}. Interestingly, beyond the anomalous Hall effect stemming from the
real-space non-coplanar spin texture, the current system can potentially have additional
contributions to the anomalous velocity stemming from the combination of spin-orbit coupling and magnetism\cite{PhysRevB.82.161414}.
It is also worth to mention that, apart from the anomalous Hall velocity for electrons, magnetic excitations
of the symmetry broken state can also show topological features. In particular, magnon excitations of the
ferromagnetic state can also potentially display anomalous Hall effect\cite{Onose2010,Owerre2016}.
At the microscopic level, this stems from the Dzyaloshinskii–Moriya interaction of the effective spin
model\cite{PhysRev.120.91,Dzyaloshinsky1958}, which naturally arises from the intrinsic Rashba
spin-orbit coupling in the electronic Hamiltonian\cite{PhysRevB.101.184401}.

\section{Exchange controlled correlations
in twisted Janus dichalcognides}
\label{sec:ex}

\begin{figure}[t!]
\centering
    \includegraphics[width=\columnwidth]{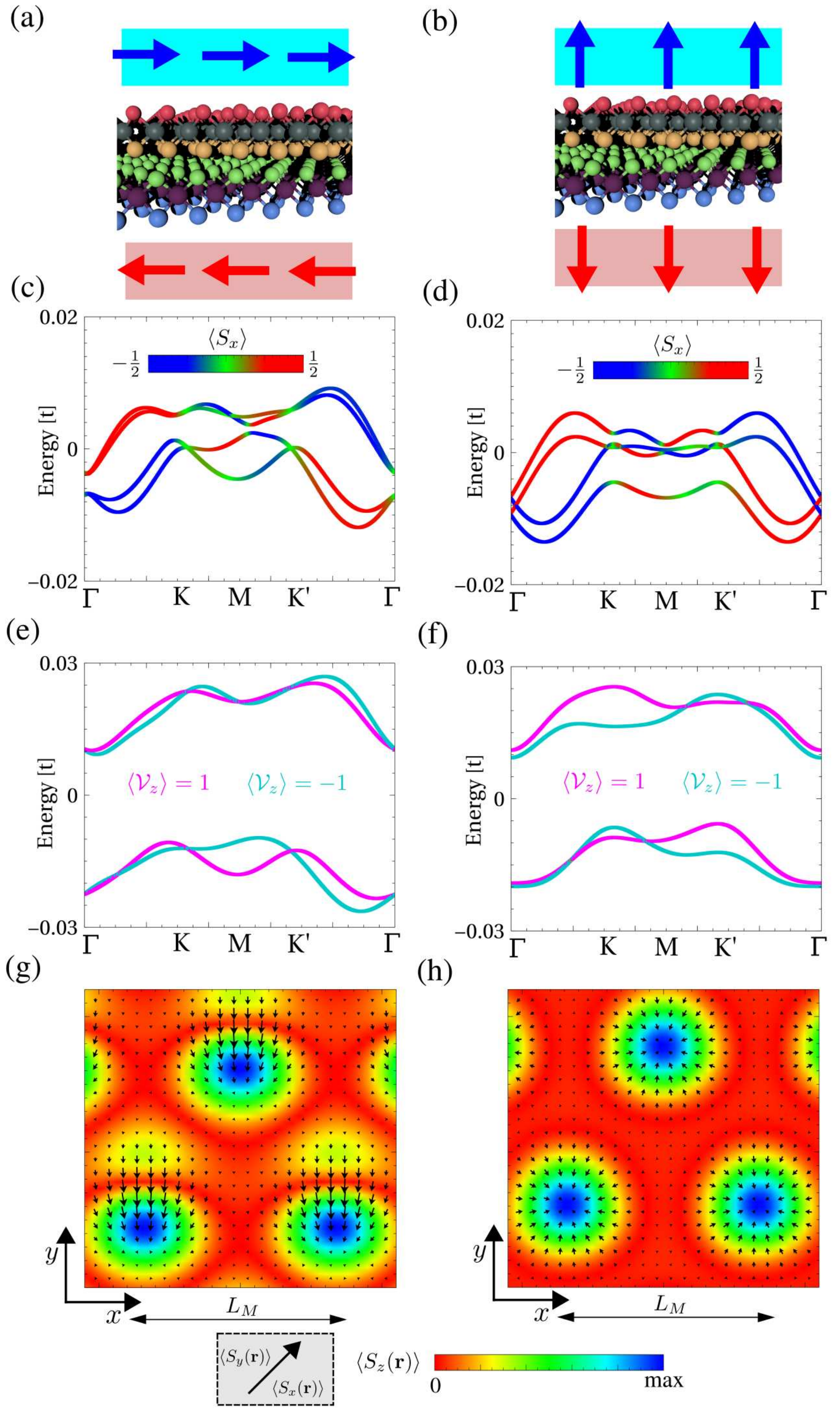}

\caption{
    Twisted Janus bilayer with in-plane (a,c,e,g)
    and out-of-plane (b,d,f,h) magnetic encapsulation.
    Panels (a,b) show a sketch of the magnetically
    encapsulated Janus bilayer,
    panels (c,d) the non-interacting band structure,
    panels (e,f) the interacting band structure
    and panels (g,h) the the real space interaction-induced
    magnetization texture.
    It is observed that the direction of the
    antiferromagnetic encapsulation impact
    both the band structure and real space
    magnetic texture distribution.
    We took $U=0.8t$, $V_1 = 0.6t$ and $V_2 = 0.2t$,
    $\lambda_I = 0.1 \lambda_R$ and $\lambda_I=0.02t$,
    and twist angle of 4.3$^\circ$.
}
\label{fig:fig5}
\end{figure}

We now move on to consider the impact of a magnetic encapsulation, and in particular, how the correlated state is impacted by it.  Twisted Janus bilayer can be encapsulated between other two-dimensional materials.  While usually boron-nitride encapsulation is used, other different insulators can be considered. In particular, the use of van der Waals ferromagnetic insulators\cite{McGuire2017,SorianoRossier2020,Blei2021} such as CrCl$_3$\cite{Wang2019}, CrBr$_3$\cite{Kim2019}, and CrI$_3$\cite{Huang2017} as encapsulation\cite{PhysRevLett.121.067701,PhysRevB.100.085128,PhysRevB.102.075435,PhysRevLett.126.056803,2021arXiv210207484C} provide an interesting possibility with regards to controlling a correlated state in the twisted Janus bilayer.  We now will show how a magnetic encapsulation allows controlling the underlying electronic structure of the Janus system, and in particular, tuning the non-coplanar magnetic order of the correlated state.  For this purpose, we now include a new term in the Hamiltonian that accounts for the impact of the effect of a magnetic substrate in the Janus bilayer, taking the form

\begin{equation}
\H_J = \sum_{i,s,s'} \tau_{ii}  \mathbf J \cdot \mathbf \sigma_{s,s'} c^\dagger_{i,s} c_{i,s'} 
\end{equation}

where $\tau_{ii}=\pm 1$ for the upper/lower layer, and $\mathbf J$ denotes the strength of the exchange field.  We focus on the case in which the ferromagnetic encapsulation is aligned antiferromagnetically, as expected from the superexchange between layers mediated by the Janus bilayer.  We will compare two scenarios for the magnetic encapsulation (Fig. \ref{fig:fig5}ab), in-plane ferromagnetics such as the case of CrCl$_3$ with $\mathbf J= (J,0,0)$, out-of-plane ferromagnets such as the case of CrBr$_3$ and CrI$_3$ with $\mathbf J= (0,0,J)$.  Let us first discuss the non-interacting electronic structure of the magnetically encapsulated bilayer show in Fig. \ref{fig:fig5}cd.  Depending on the alignment of the magnetic encapsulation, a big change in the low energy dispersion is observed, stemming from the combination of exchange proximity effect, Ising spin-orbit coupling, and Rashba spin-orbit coupling.

\begin{figure}[t!]
\centering
    \includegraphics[width=\columnwidth]{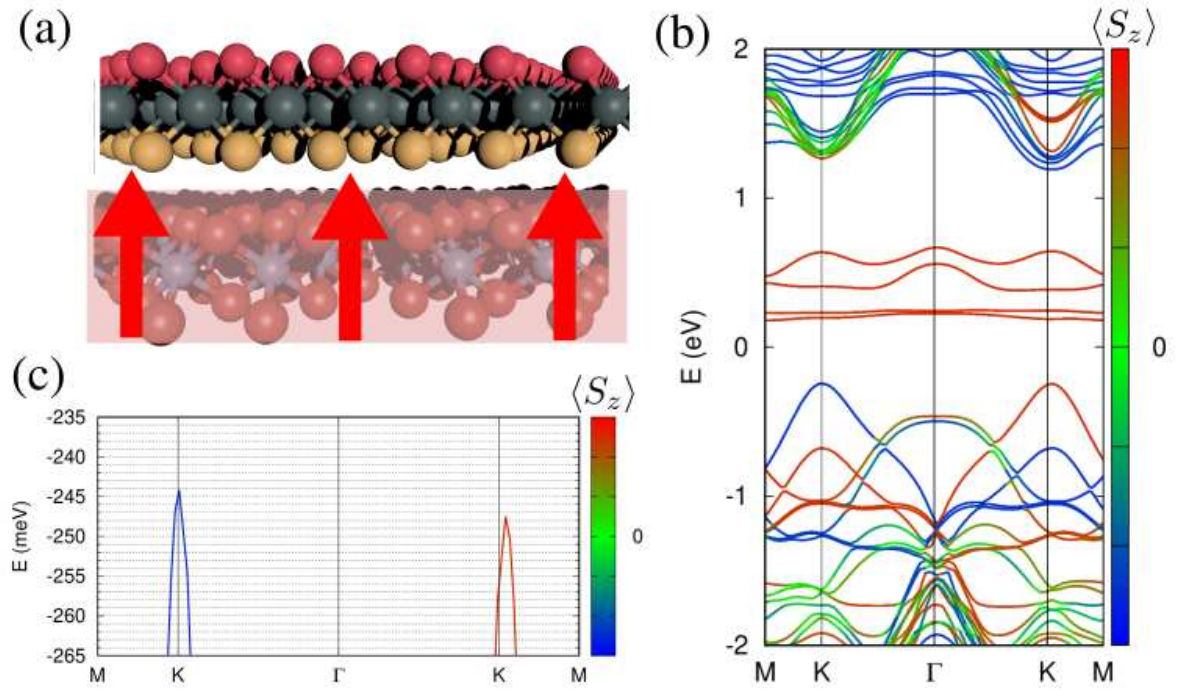}

\caption{
    Sketch of the ferromagnet/Janus
    interface with out-of-plane 
    magnetization (a).
    Panel (b) shows the
    electronic bands, highlighting
the existence of both Ising and Rashba
spin-orbit coupling.
    Panel (c) shows a zoom at the
    K-point, highlighting the exchange
    splitting induced
    by the ferromagnet at the valleys.
}
\label{fig:fig6}
\end{figure}

We now move on to consider the impact of interactions, shown in the interacting electronic structure of Fig. \ref{fig:fig5}ef.  It is interesting to note that the interaction-induced complex non-coplanar order does not impact the valley sector of the twisted dichalcogenide system.  This can be verified by computing the expectation value of the valley operator $\mathcal{V}_z$\cite{PhysRevLett.120.086603,PhysRevLett.121.146801,PhysRevB.99.245118}, defined in the real-space atomistic basis as $ \mathcal{V}_z = \frac{i}{3\sqrt{3}} \sum_{\langle \langle ij\rangle \rangle,s} \vartheta^z_{ij} \nu_{ij} c^\dagger_{i,s} c_{j,s} $.  We observe that the emergence of non-coplanar order does not create intervalley mixing, demonstrating that the complex spin texture coexists with the emergent valley conservation of the twisted dichalcogenide system. It is finally worth to emphasize that different electronic fillings can potentially lead to spontaneously valley mixed states, as proposed for twisted graphene multilayers\cite{PhysRevLett.126.056803,2021arXiv210505857K,phinney2021strong}.

More interestingly, the spatial distribution of the interaction-induced magnetic state is shown to depend strongly on the magnetic encapsulation, as shown in Fig. \ref{fig:fig5}gh. In particular, we observe that the encapsulation creates a redistribution of the direction and magnitude of the magnetic moment in the z-direction in real space. This phenomenology stems from the non-trivial interference between Wannier states of the moire unit cell, whose dependence is inherited from the underlying electronic structure featuring strong spin-orbit coupling. It is worth emphasizing that in the absence of spin-orbit coupling effects, such redistribution would not happen, and therefore this is a genuine effect arising from the interplay of spin-orbit and strong interactions.

So far we have based our analysis on an effective low energy model, in particular including the impact of the ferromagnet as an effective exchange field. We now highlight via first-principles methods that this assumption is correct a specific heterostructure, in
particular taking SWeSe as the dichalcogenide and CrBr$_3$ as the ferromagnetic insulator. We take a minimal heterostructure (Fig. \ref{fig:fig6}a) capturing the interplay of Ising spin-orbit coupling, Rashba spin-orbit coupling and exchange proximity magnetism, namely a CrBr$_3$/SWSe heterostructure. Our calculations are performed with Quantum Espresso\cite{Giannozzi2009}, including correlations in the ferromagnet by means of the DFT+U formalism and with LDA exchange-correlation functional\cite{PhysRevB.23.5048}, and PAW pseudopotentials\cite{DalCorso2014}.  As shown in Fig. \ref{fig:fig6}b, we observe that the electronic structure of the ferromagnet does not get mixed with the Janus system, highlighting that at low energies, the ferromagnet can be integrated out.  By zooming at the valleys (Fig. \ref{fig:fig6}c), we observe that the combination of spin-orbit coupling and exchange proximity effect breaks the original $+k \rightarrow -k$ degeneracy of the electronic structure, demonstrating the existence of the exchange proximity effect.  It is worth to note that, although this calculation focuses on the interplay of spin-orbit effect and exchange proximity, correlated states could be studied purely from first-principles methods\cite{2021arXiv210203259V,2020arXiv200513868Z,PhysRevB.102.235418,2020arXiv201106714P,PhysRevB.102.125424,PhysRevLett.121.266401}. These calculations would be nevertheless a remarkable challenge from the computational point of view due to the existence of non-coplanar magnetism, spin-orbit coupling, and a large number of atoms in the moire unit cell.

\section{Conclusions}
\label{sec:con}

We have shown that a twisted Janus dichalcogenide bilayer displays an electronic structure dominated by Rashba spin-orbit coupling effects.  In particular, we showed the emergence of spin-momentum spin-textures in the electronic structure, which are inherited by the flat band regime of the twisted Janus dichalcogenide system. Upon including electron-electron interactions, we showed that a non-coplanar real-space texture emerges in each unit cell of the moire system.  We further showed that the non-coplanar symmetry broken state could be controlled by magnetically encapsulating the system, providing a simple means of engineering controllable non-coplanar moire magnetism.  Our results put forward twisted Janus dichalcogenide bilayers as a versatile platform to engineer spin-orbit correlated states in moire systems, exemplifying the critical role that relativistic effects can have in flat band materials.

\textbf{Acknowledgments:}
D.S.  thanks financial support from EU through the MSCA project Nr. 796795
SOT-2DvdW.  J.L.L. acknowledges the computational resources provided by the
Aalto Science-IT project, and the financial support from Academy of Finland  Projects No. 331342 and No. 336243.
We thank P. Liljeroth, S. Kezilebieke, T. Heikkil\"{a}, F. Aikebaier, T. Wolf, P. T\"{o}rm\"{a}
and A. Ramires for fruitful discussions.  

\bibliography{biblio}{}
\end{document}